\newcommand{\eq}[1]{\begin{align}#1\end{align}}

\documentclass[twocolumn,aps,prb,superscriptaddress]{revtex4-1}
\usepackage{bm}
\usepackage{amsmath,amsthm,amssymb}
\usepackage{color}
\usepackage{newtxmath}
\usepackage{array,booktabs}
\usepackage[dvipdfmx]{graphicx}
\usepackage{here}

\begin{document}
\title{Spin-Dependent Conductance in a Junction with Dresselhaus Spin-Orbit Coupling}

\author{Daisuke Oshima}
\affiliation{Department of Applied Physics, Nagoya University, Nagoya, 464-8603, Japan}

\author{Katsuhisa Taguchi}
\affiliation{Yukawa Institute for Theoretical Physics, Kyoto University, Kyoto, 606-8502, Japan}

\author{Yukio Tanaka}
\affiliation{Department of Applied Physics, Nagoya University, Nagoya, 464-8603, Japan}

\begin{abstract}
We studied spin-dependent conductance in a normal metal (NM)/NM junction with Dresselhaus spin-orbit coupling (DSOC) and magnetization along the out-of-plane direction. As a reference, we also studied the spin-dependent conductance in such a junction with 
Rashba spin-orbit coupling (RSOC). 
Using a standard scattering method, we calculated the gate-voltage dependence of the spin-dependent conductances in DSOC and RSOC.
In addition, we calculated the gate-voltage dependence of the conductances in a ferromagnetic metal (FM)/NM junction with spin-orbit coupling and magnetization, which we call ferromagnetic spin-orbit metal (FSOM). 
From these results, we discuss the relation between these conductance in the presence of DSOC and that in the presence of RSOC.
We found that conductance in DSOC is the same as that in RSOC for the NM/FSOM junction.
In addition, we found that in the FM/FSOM junction, the conductance in DSOC is the same as that in RSOC. 
\end{abstract}

\maketitle
\section{Introduction}
Spin-dependent transport is a key issue in the context of spintronics. 
Recently, various spin-dependent transports in the presence of spin-orbit couplings (SOCs) have been discussed.
Among the spin-dependent transports due to the SOCs, Rashba and Dresselhaus spin-orbit coupling (RSOC and DSOC, respectively) have been intriguingly studied\cite{Grundler01,Larsen02,Streda03,Perel03,Krupin05,Wang05,Sanchez08,Lucignano08,Sakr11,Fallahi12,Pang12,Tang12,Jantayod13,Jantayod15,Fukumoto15,Han15,Wojcik15,Tang16,Tang17,Oshima18}. 
RSOC and DSOC are caused by the structural and bulk inversion symmetry breaking, respectively\cite{Rashba60,Dresselhaus55,Ganichev14,Dario15,Kohda17}.

\begin{figure}[h]
\centering
\includegraphics[width = 85mm]{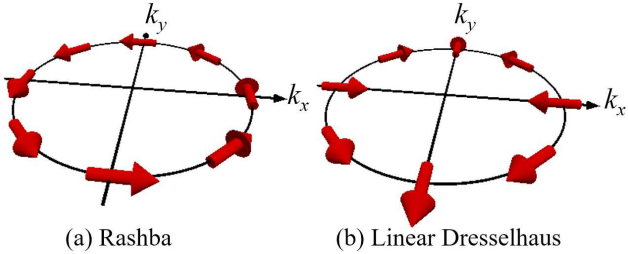}
\caption{(Color online) 
Illustration of the spin texture (red arrows) of the (a) Rashba and (b) linear Dresselhaus-type spin-orbit coupling.
}
\label{magneticfied}
\end{figure}

In the context of charge transport in a diffusive regime, the transport as well as electromagnetic effects (e.g., Edelstein effect) in the presence of RSOC have been studied so far\cite{Taguchi17} and it depends on the spin textures:
The spin texture at the Fermi surface in the momentum space, as shown in Fig. \ref{magneticfied},  is shifted by the applied electric field; this shift is proportional to the electric field and the transport relaxation time of the diffusive regime.
Hence, the spin polarization is generated by the applied electric field and its polarization is perpendicular (parallel) to the applied electric field in the presence of RSOC (DSOC). 
%
%

In the context of a ballistic regime, spin-dependent tunneling conductance has been studied in two-dimensional systems\cite{Molenkamp01,Grundler01,Matsuyama02,Larsen02,Jiang03,Ramaglia03,Streda03,Krupin05,Yokoyama06,Srisongmuang08,Sanchez08,Lucignano08,Matos-Abiague09,Fallahi12,Pang12,Tang12,Zhang13,Jantayod13,Jantayod15,Fukumoto15,Han15,Wojcik15,Tang16,Oshima18}. 
For example, the spin-dependent conductance (e.g., magnetoresistance) depends on the RSOC spin texture, which is discussed in the surface state of a three-dimensional topological insulator with magnetism. 
However, the charge conductivity is not largely influenced by the spin textures, for example, in a normal metal (NM)/NM junction with RSOC and magnetization 
\cite{Oshima18}.
In this system, although such an NM with RSOC and magnetism has three types of spin textures depending on the Fermi level, 
the charge conductance could be independent of the spin texture of the RSOC.

%
\begin{figure}[b]
\centering
\includegraphics[width = 85mm]{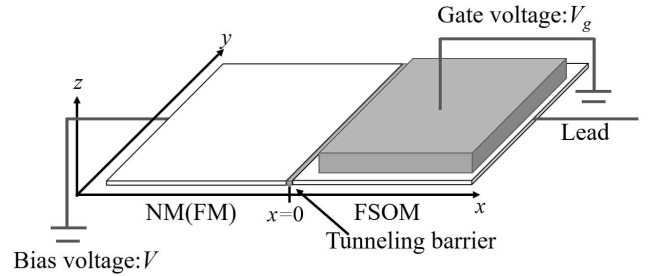}
\caption{
Illustration of a two-dimensional NM (FM)/FSOM junction, where the FSOM has RSOC or DSOC and magnetization.
The magnetization is along the out-of-plane direction. 
The Fermi level as well as the spin texture of the FSOM is changed by an applied gate voltage $V_g$. 
A bias voltage ($V$) is applied on the NM (FM).}
\label{junctionmodel}
\end{figure}
%
It is a natural to ask whether the charge conductance is independent of the structure of the spin texture and how the spin-dependent conductance depends on the spin texture in the ballistic regime in the presence of DSOC. 
In this work, in order to address this question, we studied the relation between the conductance and spin texture in an NM/NM junction with DSOC and magnetism. 
In this paper, for simplicity, 
we call an NM with SOC and magnetism a ferromagnetic spin-orbit metal (FSOM). We calculated the charge and spin-dependent conductance when the spin textures of the FSOM depend on the Fermi level, which is tuned by an applied gate voltage in the NM/FSOM junction, as shown in Fig. \ref{junctionmodel}.
As a result, we show the relation between these conductances in the DSOC and those in the RSOC.
Furthermore, the spin-dependent conductance is discussed in a ferromagnetic metal (FM)/FSOM junction. 
We found that the charge conductance in the DSOC is the same as that in the RSOC for the NM/FSOM junction.
The spin-dependent conductance in the DSOC is different from that in the RSOC, 
but there are some relations between them.

The organization of this paper is as follows. 
In Sect. \ref{sec:II}, we describe the model and our obtained results.
In Sect. \ref{sec:III}, we discussed the results through the transformation.
In Sect. \ref{sec:IV}, we summarize the obtained results.

\section{Model and Results}\label{sec:II}
\subsection{Model}
We first consider a two-dimensional NM/FSOM junction as illustrated in Fig. \ref{junctionmodel}, 
where the RSOC or DSOC of the FSOM is induced by the structural or bulk inversion symmetry breaking.
A bias voltage $V$ (to drive charge along the $x$ direction) and a gate voltage $V_g$ (to change the Fermi level in the FSOM\cite{Datta90}) are applied on the NM and FSOM, respectively.
The tunneling barrier at the interface is assumed to be a delta function\cite{Srisongmuang08,Jantayod13,Fukumoto15} for conservation of the $y$ component of the momentum.
We also assume the case where the width of the junction along the $y$ direction is sufficiently large.

The model Hamiltonian of the NM/FSOM junction can be described as \cite{Rashba60,Cayao15,Streda03,Fukumoto15,Oshima18}  
\begin{align}
\label{eq:total}
&H= H_{L}\theta(-x) + U\delta(x) + H_{R}\theta(x), \\
&H_{L} = \frac{\hbar^2k^2}{2m_L},\nonumber\\
&H_{R} = \frac{\hbar^2k^2}{2m_R}+{H}_{{\rm SOC}}-M_R\sigma_z+eV_g,\nonumber
\end{align}
where $H_{L}$ ($H_{R}$) is the Hamiltonian in the NM (FSOM). 
Here, $\theta$ and $\delta$ are the Heaviside step function and delta function, respectively. 
The first term in Eq. (\ref{eq:total}) denotes the kinetic energy, $k^2=k^2_x+k^2_y$, and $m_L(m_R)$ is the effective mass of the electron in the NM (FSOM).
The second term in $H_{\rm R}$ denotes the SOCs\cite{Ganichev14,Kohda17}:
\begin{align}
H_{\rm SOC}=\left\{
\begin{aligned}
&\alpha(k_x\sigma_y-k_y\sigma_x) & \qquad(\textrm{RSOC}),\\
&\beta(k_x\sigma_x-k_y\sigma_y) + o(k^3) &\qquad(\textrm{DSOC}),
\end{aligned}
\right.
\end{align}
where ${\bm \sigma}=(\sigma_x,\sigma_y,\sigma_z)$ are Pauli matrices in the spin space.
$\alpha$ and $\beta$ are a coupling constant of the RSOC and DSOC, respectively. 
Here, we consider the linear DSOC, neglecting a term proportional to $k^3$.
The third term in $H_{\rm R}$, $M_R\sigma_z$, denotes the exchange coupling of the magnetization, which is along the out-of-plane direction.
$eV_g(\geq 0)$ is a potential caused by the gate voltage.
The third term in Eq. (\ref{eq:total}), $U\delta(x)$, indicates the tunneling barrier, where $U$ is the strength of the tunneling barrier.

\subsection{Transmission probability in an NM/FSOM junction}\label{NM/FSOM}
Using a standard method, we calculated the angle-resolved transmission probability $T^{\gamma,s}_{A=R,D}(\phi)$,
where $\phi$ is an injection angle from the NM into the FSOM; it is defined as an angle between the wave vector of the injection wave and the $x$-axis.
Here, the superscript $s(=\uparrow,\downarrow$) denotes the up- and down-spin along the spin polarization $\gamma (=x,y,z)$ in the NM. The subscript $A(=R,D)$ indicates the SOC (RSOC or DSOC). 
The detail of the derivation of $T^{\gamma,s}_{A}$ is shown in the Appendix \ref{app1}.
The transmission probability 
$\mathcal{T}^{\gamma,s=\uparrow,\downarrow}_{A}$ is given by 
\begin{align}
\mathcal{T}^{\gamma,s=\uparrow,\downarrow}_{A}=\frac{1}{2\pi}\int^{\frac{\pi}{2}}_{-\frac{\pi}{2}}d\phi k_F \cos\phi \cdot T^{\gamma,s}_{A}(\phi),
\label{def:T}
\end{align}
where $k_F=\sqrt{2mE_F}/\hbar$ is the Fermi momentum in the NM.
$\mathcal{T}^{\gamma,s}_{A}$ and $\mathcal{T}^{\gamma,\uparrow}_{A}+\mathcal{T}^{\gamma,\downarrow}_{A}$ are proportional to the spin-dependent conductance and the charge conductance, respectively (see Appendix \ref{app1}). 
It is found that at $\alpha = \beta$, the energy dispersion of the FSOM of the 
RSOC is the same as the energy dispersion of the DSOC, but the spin texture of 
the RSOC is different from that of the DSOC (see Fig. \ref{magneticfied}). 
In order to show the relation between the conductances (i.e. transmission probabilities) and these spin textures, we simply set $\alpha=\beta >0$, $M_R \ge0$, and $m_L=m_R\equiv m$.

\begin{figure}[htbp]
\centering
\includegraphics[width = 85mm]{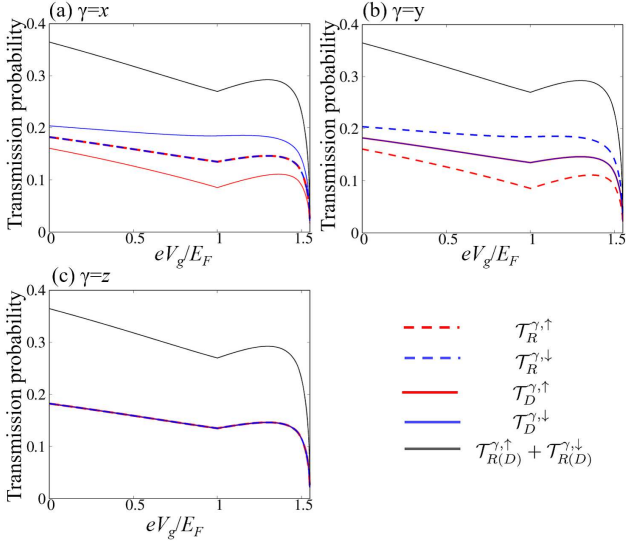}
\caption{(Color online) $V_g$ dependence of the transmission probability of the RSOC $\mathcal{T}^{\gamma,s}_{R}$ (dashed lines) and that of the DSOC $\mathcal{T}^{\gamma,s}_{D}$ (bold lines) in the NM/FSOM junction with several spin 
polarizations of the NM: (a) $\gamma=x$, (b) $\gamma=y$, and (c) $\gamma=z$ at $M_R/E_F=0$.  
Red and blue lines correspond to the up- and down-spin polarized electron 
injection case from the NM into the FSOM, respectively.
Black lines show the total transmission probability. 
Here, we set $E_\alpha/E_F=0.55$, $Uk_F/E_F=1.0$, and $\alpha =\beta$, where $E_\alpha=m\alpha^2/(2\hbar^2)$ is the Rashba energy\cite{Ast07,Sablikov07,Srisongmuang08,Ast08,Mathias10,Ishizaka11,Jantayod13,Cayao15}.}
\label{result1}
\end{figure}
%
Fig. \ref{result1} shows the $V_g$ dependence of 
$\mathcal{T}^{\gamma,s}_{A=R,D}$ for $M_R=0$ with several spin polarizations of 
the NM $\gamma=x,y,z$. 
It is found that at $\gamma=x$, the $V_g$ dependence of $\mathcal{T}^{x,s}_{R}$ is distinct at the kink $eV_g/E_F=1$; 
for $eV_g/E_F\geq1 $, the inner and outer spin textures are pointing along the 
counterclockwise direction in the momentum space, whose spin texture is caused by 
the RSOC and the two spin-split bands; 
for $eV_g/E_F\leq1 $, the inner and outer spin textures are pointing oppositely. 
The $\mathcal{T}^{x,s}_{R}$ is independent of $s(=\uparrow, \downarrow)$ of the FSOM. 
On the other hand, the $V_g$ dependence of $\mathcal{T}^{x,s}_{D}$ depends on $s$: 
the $\mathcal{T}^{x,\uparrow}_{D}$ clearly has the kink at $eV_g/E_F=1$, but the $\mathcal{T}^{x,\downarrow}_{D}$ does not. 
Figure \ref{result1}(b) indicates the $V_g$ dependence of the $\mathcal{T}^{y,s}_{A=R,D}$ for $\gamma=y$.
Then, it is noted that the $\mathcal{T}^{y,s}_{R}$ depends on $s$, but $\mathcal{T}^{y,s}_{D}$ is independent of $s$.
For $\gamma=z$, both $\mathcal{T}^{z,s}_{R}$ and $\mathcal{T}^{z,s}_{D}$ are equal and independent of $s$.
Moreover, it is found that for all $\gamma$, 
the total transition probability $\mathcal{T}^{\gamma,\uparrow}_{R}+\mathcal{T}^{\gamma,\downarrow}_{R}$ in the RSOC case, which is proportional to the charge conductance, is equal to that in the DSOC.
Then, it is independent of $\gamma$.

From these numerical results, it is found that the transmission probability for any $\gamma$ satisfies the following relation: 
\begin{align}
\mathcal{T}^{\gamma, \uparrow}_{R} + \mathcal{T}^{\gamma,\downarrow}_{R} =\mathcal{T}^{\gamma, \uparrow}_{D} + \mathcal{T}^{\gamma,\downarrow}_{D}.
\label{eq:sum}
\end{align}
Furthermore, we note that the following relation between $\mathcal{T}^{\gamma,s}_{R}$ and $\mathcal{T}^{\gamma,s}_{D}$ is satisfied for any $V_g$:
\begin{align}
\begin{aligned}
&\mathcal{T}^{x,\uparrow (\downarrow)}_D=\mathcal{T}^{y,\uparrow (\downarrow)}_R,\\
&\mathcal{T}^{x,\uparrow}_R=\mathcal{T}^{x,\downarrow}_R=\mathcal{T}^{y,\uparrow}_D=\mathcal{T}^{y,\downarrow}_D,\\
&\mathcal{T}^{z,\uparrow}_R=\mathcal{T}^{z,\downarrow}_R=\mathcal{T}^{z,\uparrow}_D=\mathcal{T}^{z,\downarrow}_D.
\end{aligned}
\label{eq:result1}
\end{align}
Thus, there are some relations of the transmission probabilities without the 
magnetization case ($M_R=0$).

\begin{figure}[htbp]
\centering
\includegraphics[width = 85mm]{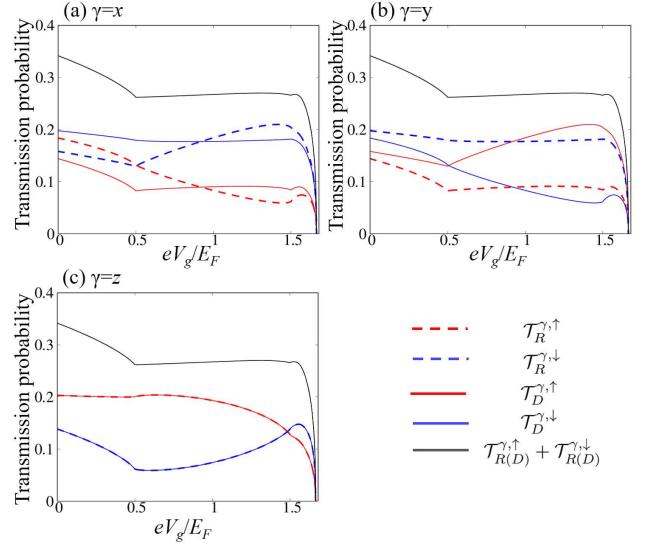}
\caption{(Color online) $V_g$ dependence of $\mathcal{T}^{\gamma,s}_{A(=R,D)}$ in the NM/FSOM junction for $M_R/E_F=0.5$.
(a) $\gamma=x$, (b) $\gamma=y$, and (c) $\gamma=z$ at $E_\alpha/E_F=0.55$, $Uk_F/E_F=1.0$, and $\alpha=\beta$. }
\label{result2}
\end{figure}
%
Next, we consider $\mathcal{T}^{\gamma,s}_{A}$ in the presence of 
$M_R\neq 0$. 
Figure \ref{result2} shows the $V_g$ dependence of $\mathcal{T}^{\gamma,s}_{A}$ for nonzero $M_R$.
In particular, under the RSOC with nonzero $M_R$, 
there are two spin-split bands and three remarkable electric states, the so-called normal Rashba metal (NRM), anomalous Rashba metal (ARM)\cite{Fukumoto15}, and Rashba ring metal (RRM)\cite{Oshima18}.
In the presence of DSOC with nonzero $M_R$, we can find three similar states.
As a result, the $V_g$ dependence of the probability 
$\mathcal{T}^{\gamma,s}_{A}$ has two kinks at $eV_g/E_F = 0.5$ and 1.5, whose kinks indicate the boundary between the states.
For $\gamma=y$, the probability $\mathcal{T}^{\gamma,s}_{R}$ for $M_R\neq 0$ depends on $s$ as well as that for $M_R=0$.  
The magnitude of the probability for up-spin $\mathcal{T}^{y,\uparrow}_{R}$ is smaller than that for the down-spin $\mathcal{T}^{y,\downarrow}_{R}$ for any $V_g$. 
In other words, under nonzero $M_R$, the difference between the probabilities $\mathcal{T}^{\gamma,\uparrow}_{R} - \mathcal{T}^{\gamma,\downarrow}_{R} $ 
takes a negative value.
On the other hand, for $\gamma=x,z$, the probability $\mathcal{T}^{\gamma,s}_{R}$ for $M_R\neq 0$ depends on $s$ and complicatedly depends on $V_g$, unlike that for $M_R=0$.
In $\gamma=x$, the difference between the probabilities takes a positive value in the NRM and a negative value in the ARM and RRM.
In $\gamma=z$, the difference takes a positive value in the NRM and ARM but it becomes a negative value in the RRM.
Note that the $V_g$ dependence of the sign of the difference depends on the parameters, and it does not completely correspond to the three states.

We also numerically calculated $\mathcal{T}^{\gamma,s}_{D}$ in $M_R\neq 0$, as shown in Fig. \ref{result2}. 
As a result, we found the following relation between $\mathcal{T}^{\gamma,s}_{D}$ and $\mathcal{T}^{\gamma,s}_{R}$ for any $V_g$: 
\begin{align}
\mathcal{T}^{x,s}_D=\mathcal{T}^{y,s}_R,\quad\mathcal{T}^{x,\uparrow(\downarrow)}_R=\mathcal{T}^{y,\downarrow(\uparrow)}_D,\quad \mathcal{T}^{z,s}_R=\mathcal{T}^{z,s}_D.
\label{eq:result2}
\end{align}
Furthermore, the total transition probability is independent of $\gamma$: Eq. (\ref{eq:sum}) is satisfied even for nonzero $M_R$.

\subsection{Angle-resolved transmission probability}\label{angle_NM/FSOM}
In order to show the SOC dependence of the probability in more detail, we numerically calculated the angle-resolved transmission probability $T^{\gamma,s}_{A(=R,D)}(\phi)$ defined in Eq. (\ref{def:T}). 
%
\begin{figure}[htbp]
\centering
\includegraphics[width = 85mm]{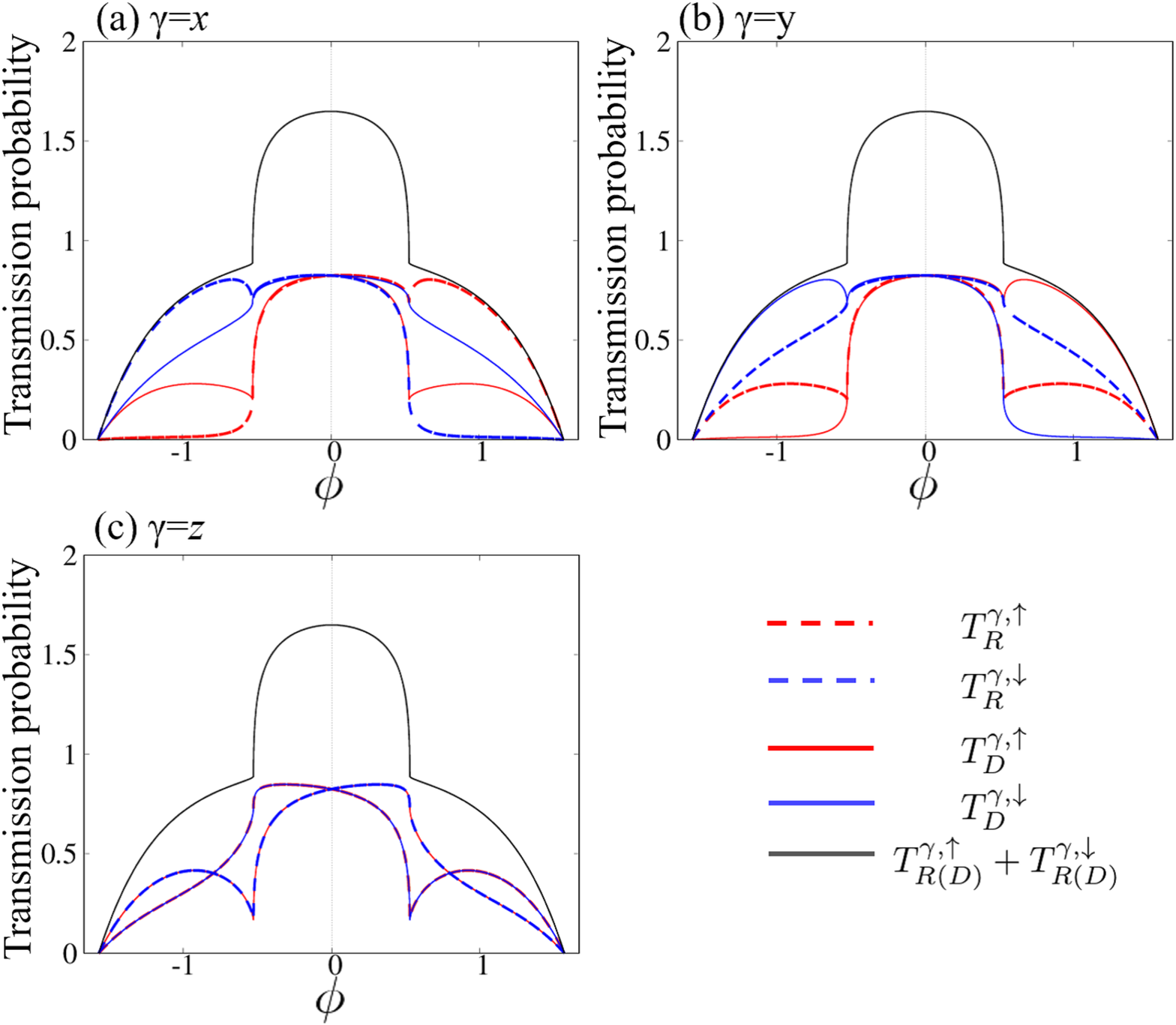}
\caption{(Color online) $\phi$ dependence of $T^{\gamma,s}_{A(=R,D)}$ in the NM/FSOM junction for $M_R/E_F=0$:
(a) $\gamma=x$, (b) $\gamma=y$, and (c) $\gamma=z$ at $E_\alpha/E_F=0.55$, $Uk_F/E_F=1.0$, $eV_g/E_F=0$, and $\alpha=\beta$.  
}
\label{result3}
\end{figure}
%
Figure \ref{result3} shows the $\phi$ dependence of $T^{\gamma,s}_{R}(\phi)$ for $M_R/E_F=0$ and $eV_g/E_F=0$, i.e., in the NRM. 
The angle-resolved probability $T^{\gamma,s}_{R}(\phi)$ complicatedly depends on $\phi$. 
However, $T^{\gamma,s}_{R}(\phi)$ implies a symmetric to $\phi$ and $s$:
\begin{align}
\begin{aligned}
&T^{x,\uparrow}_R(\phi)=T^{x,\downarrow}_R(-\phi),\\
&T^{y, s}_R(\phi)=T^{y,s}_R(-\phi),\\
&T^{z,\uparrow}_{R}(\phi)=T^{z,\downarrow}_{R}(-\phi).
\end{aligned}
\label{eq:result3.1}
\end{align}
Thus, these relations are categorized by $\gamma=x,y, z$. 
We also found the relation between $T^{\gamma,s}_{R}(\phi)$ and $T^{\gamma,s}_{D}(\phi)$. 
The relations are given by
\begin{align}
T^{x(y),s}_{R}(\phi)=T^{y(x),s}_{D}(\phi),\quad T^{z,\uparrow(\downarrow)}_{R}(\phi)=T^{z,\downarrow(\uparrow)}_{D}(\phi).
\label{eq:result3.2}
\end{align} 
Furthermore, it is found that in the NRM, the total probability for each angle is given by 
\begin{align}
T^{\gamma, \uparrow}_{R}(\phi) + T^{\gamma,\downarrow}_{R} (\phi) = T^{\gamma, \uparrow}_{D}(\phi) + T^{\gamma,\downarrow}_{D}(\phi).
\end{align}
We also confirmed these relations even in the RRM. 

%
\begin{figure}[htbp]
\centering
\includegraphics[width = 85mm]{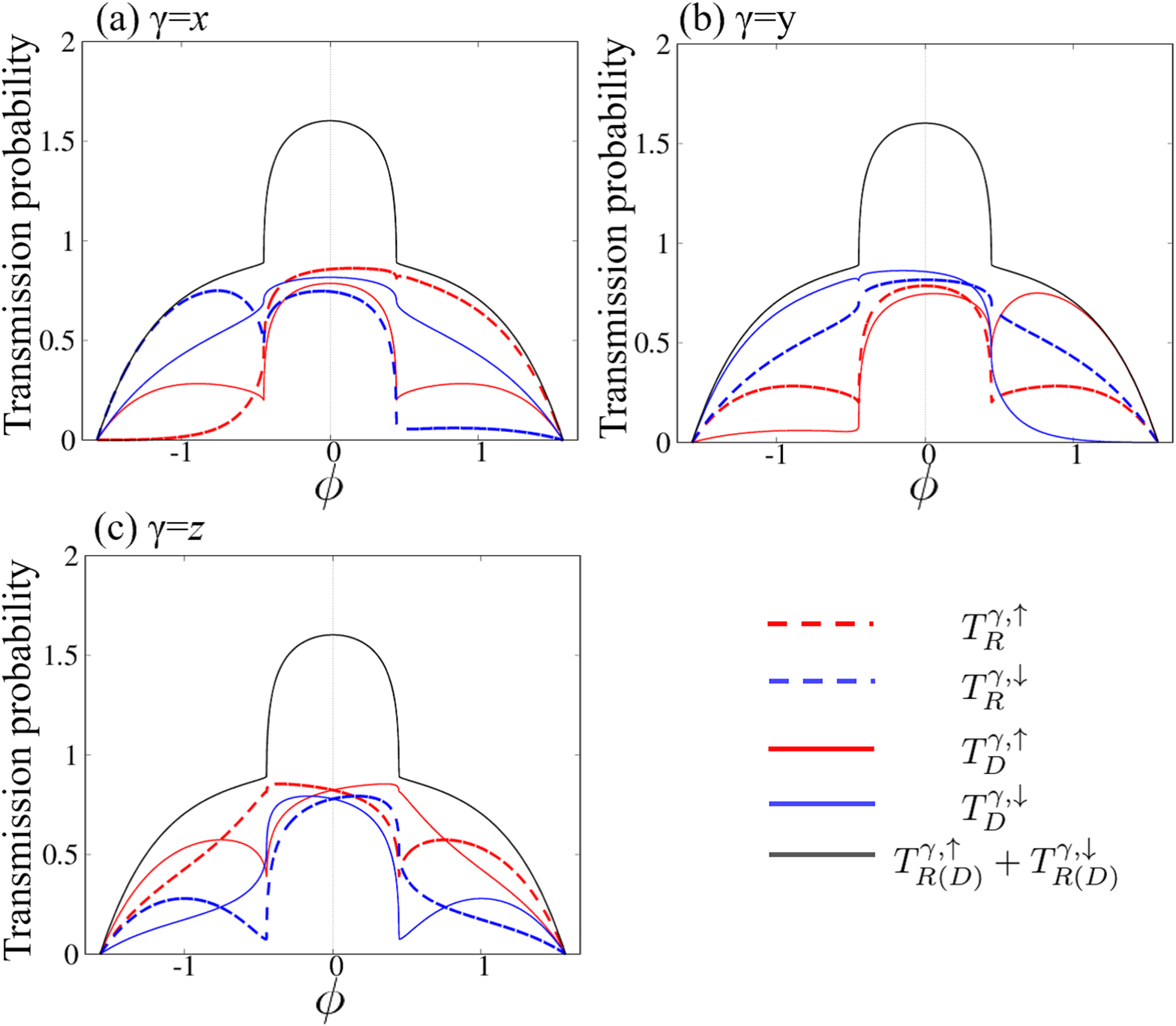}
\caption{(Color online) $\phi$ dependence of $T^{\gamma,s}_{A(=R,D)}$ in the NM/FSOM junction for $M_R/E_F=0.5$:
(a) $\gamma=x$, (b) $\gamma=y$, and (c) $\gamma=z$.
Here, we set $E_\alpha/E_F=0.55$, $Uk_F/E_F=1.0$, $eV_g/E_F=0$, and $\alpha=\beta$. 
}
\label{result4}
\end{figure}
%
The angle-resolved probability under nonzero $M_R$ is shown in Fig. \ref{result4}. 
The $\phi$ dependence of the angle-resolved transmission probability is highly complicated. 
However, we found that only in $\gamma=y$, the probability $T^{y,s}_R(\phi)$ is symmetric for $\phi \to -\phi$:
\begin{align}
T^{y,s}_R(\phi)=T^{y,s}_R(-\phi), 
\label{eq:result4.1}
\end{align}
The probability of the RSOC and that of the DSOC under $M_R\neq0$ 
satisfies the following relations: 
\begin{align}
\begin{aligned}
&T^{x,s}_D(\phi)=T^{y,s}_R(\phi),\\
&T^{y,\downarrow(\uparrow)}_D(\phi) = T^{x,\uparrow(\downarrow)}_R(-\phi),\\
&T^{z, s}_{D}(\phi)=T^{z, s}_{R}(-\phi).
\end{aligned}
\label{eq:result4.2}
\end{align}
Thus, for nonzero $M_R$, $T^{x,s}_{R}(\phi)$ is equal to $T^{y,s}_{D}(\phi)$ by $\phi\to -\phi$ and $ s=[\uparrow (\downarrow)] \to [\downarrow (\uparrow)] $, and $T^{z,s}_{R}(\phi)$ is equal to $T^{z,s}_{D}(\phi)$ by only $\phi\to -\phi$.
It is noted that these relations are also numerically confirmed even in 
the case of ARM and RRM.

\subsection{Transmission probability in an FM/FSOM junction}\label{FM/FSOM}
In the previous subsection \ref{NM/FSOM} and \ref{angle_NM/FSOM}, we show the relation between $\mathcal{T}^{\gamma, s}_{R}$ and $\mathcal{T}^{\gamma, s}_{D}$ in the NM/FSOM junction.
For $M_R\neq 0$, the FSOM of the NM/FSOM junction breaks a time-reversal symmetry.  
Next, we consider an FM/FSOM junction without time-reversal symmetry in the whole of the junction. 
We studied the probability in the junction with the lower symmetry.
The model of the FM/FSOM junction can be described as
\begin{align}
\label{eq:FMFSOM}
&H= H_{L}\theta(-x) + U\delta(x) + H_{R}\theta(x) , \\
&H_{L} = \frac{\hbar^2k^2}{2m_L}-{\bm M}_L\cdot{\bm \sigma},\nonumber\\
&H_{R} = \frac{\hbar^2k^2}{2m_R}+{H}_{{\rm SOC}}-M_R\sigma_z+eV_g,\nonumber
\end{align}
where ${\bm M}_L$ denotes the magnetization of the FM and ${\bm M}_L/M_L$ corresponds to $\gamma$ as defined in the previous section. 
Then, the transition probability $\mathcal{T}^{\gamma, s}_{A}$ is given by
\begin{align}
\mathcal{T}^{\gamma, s}_{A}=\frac{1}{2\pi}\int^{\frac{\pi}{2}}_{-\frac{\pi}{2}}d\phi k_F^{s} \cos\phi \cdot T^{\gamma, s}_A(\phi),
\end{align}
where $k^{s= \uparrow(\downarrow)}_F=\sqrt{2m[E_F+(-) M_L]}/\hbar$ is the momentum of up- (down-) spin electron in the FM at the Fermi level. 
Herein, we simply set $m_L=m_R\equiv m$. 

\begin{figure}[htbp]
\centering
\includegraphics[width = 85mm]{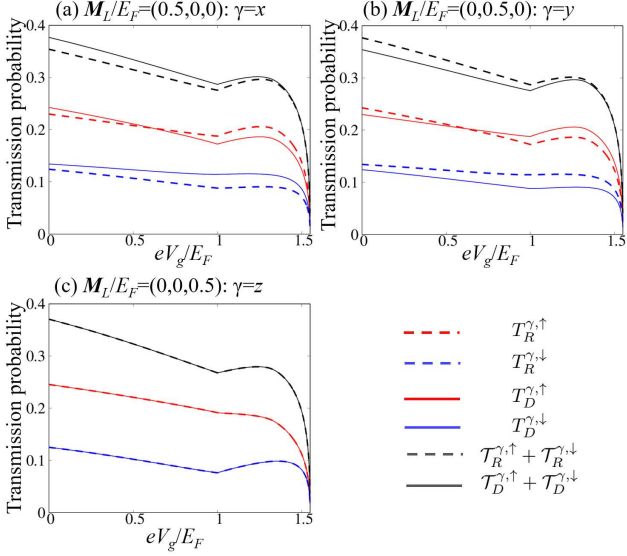}
\caption{(Color online) $V_g$ dependence of $\mathcal{T}^{\gamma, s}_{A(=R,D)}$ in the FM/FSOM junction for $M_R/E_F=0$:
(a) ${\bm M}_L/E_F=(0.5,0,0)$, (b) ${\bm M}_L/E_F=(0,0.5,0)$, and (c) ${\bm M}_L/E_F=(0,0,0.5)$ correspond to $\gamma=x$, $\gamma=y$, and $\gamma=z$, respectively.
Here, we set $E_\alpha/E_F=0.55$, $Uk_F/E_F=1.0$, and $\alpha=\beta$ with $k_F=\sqrt{2mE_F}/\hbar$.
}
\label{result5}
\end{figure}
%
Figure \ref{result5} shows the $V_g$ dependence of $\mathcal{T}^{\gamma, s}_{A}$ in $M_L\neq 0$ and $M_R=0$.
The $V_g$ dependence of $\mathcal{T}^{\gamma, s}_{R}$ as well as $\mathcal{T}^{\gamma, s}_{D}$ is similar to that in Fig. 3. 
However, in contrast to Fig. 3, $\mathcal{T}^{\gamma, s}_{R}$ and $\mathcal{T}^{\gamma, s}_{D}$ also depends on $s$ in all $\gamma$.
Furthermore, for $\gamma=z$, the relation $\mathcal{T}^{z, s}_{R}=\mathcal{T}^{z, s}_{D}$, which is satisfied for $M_L=0$, remains even in $M_L\neq0$. 
From these numerical results, 
the summary of the relation among $\mathcal{T}^{\gamma, s}_{R}$ ($\mathcal{T}^{\gamma, s}_{D}$), $s$, and $\gamma$ for any $V_g$ is given by  
\begin{align}
\mathcal{T}^{z,\uparrow}_R=\mathcal{T}^{z,\downarrow}_R, \quad \mathcal{T}^{z,\uparrow}_D=\mathcal{T}^{z,\downarrow}_D.
\label{eq:result5.1}
\end{align}
In the FM/FSOM junction with $M_R=0$, we find 
\begin{align}
\mathcal{T}^{x, s}_D=\mathcal{T}^{y, s}_R,\quad \mathcal{T}^{z,s}_D = \mathcal{T}^{z, s}_R.
\label{eq:result5.2}
\end{align}
The total probability has a relation only for $\gamma=z$:
\begin{align}
\mathcal{T}^{z, \uparrow}_{R} + \mathcal{T}^{z,\downarrow}_{R} = \mathcal{T}^{z, \uparrow}_{D} + \mathcal{T}^{z,\downarrow}_{D}.
\label{eq:result5.3}
\end{align}

\begin{figure}[htbp]
\centering
\includegraphics[width = 85mm]{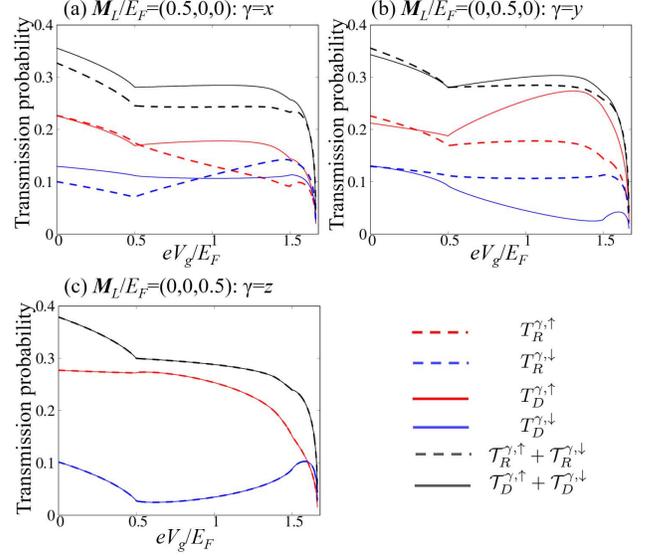}
\caption{(Color online) $V_g$ dependence of $\mathcal{T}^{\gamma, s}_{A(=R,D)}$ in the FM/FSOM junction for $M_R/E_F=0.5$:
(a) ${\bm M}_L/E_F=(0.5,0,0)$, (b) ${\bm M}_L/E_F=(0,0.5,0)$, and (c) ${\bm M}_L/E_F=(0,0,0.5)$.
Here, we set $E_\alpha/E_F=0.55$, $Uk_F/E_F=1.0$, and $\alpha=\beta$.  
}
\label{result6}
\end{figure}
%
We also numerically calculated the probability for $M_L\neq 0$ and $M_R\neq 0$, as shown in Fig. \ref{result6} (cf. Fig. \ref{result2}). 
As a result, we found that the transmission probability has a unique $V_g$ dependence: 
In $\gamma=x$, the $V_g$ dependence of $\mathcal{T}^{\gamma, s}_{A}$ has a kink at $eV_g/E_F=0.5$ and $1.5$, whose properties are similar to those in Fig. \ref{result2}. 
However, the $V_g$ dependence of the probability around the RRM in Fig. \ref{result6} monotonically decrease with increasing $eV_g/E_F$, unlike that in Fig. \ref{result2}.  
In particular, in the RSOC, the kink between NRM and ARM is clear but that between ARM and RRM disappears near $eV_g/E_F=1.5$. 
For $\gamma=y$, the $V_g$ dependence of $\mathcal{T}^{\gamma, s}_{A}$ is similar to that in Fig. \ref{result2}. 
For $\gamma=z$, we found that 
$\mathcal{T}^{\gamma, s}_{R}$ and $\mathcal{T}^{\gamma, s}_{D}$ are equal.
Then, Eq. (\ref{eq:result5.3}) is satisfied even for nonzero $M_L$ and $M_R$ only when $\gamma=z$.
However, in the FM/FSOM junction with $M_R\neq 0$, we cannot find the relation between $\mathcal{T}^{\gamma, s}_{R}$ and $\mathcal{T}^{\gamma, s}_{D}$, 
unlike in the case of an FM/FSOM junction with $M_R=0$.

In Figs. \ref{result5} and \ref{result6}, we can find that the sign of the difference $\mathcal{T}^{x(y), \uparrow}_{A}-\mathcal{T}^{x(y), \downarrow}_{A}$ 
is opposite to that in the NM/FSOM junction (see Figs. \ref{result1} and \ref{result2}).
We can numerically confirm that this is caused by a strong $M_L>0$.
For small $M_L$, the sign of the difference is the same as that in Figs. \ref{result1} and \ref{result2}.
For sufficiently large $M_L>0$, the sign switches oppositely.
However, the $V_g$ dependence of each $\mathcal{T}^{\gamma, s}_{A}$ in the FM/FSOM junction is similar to that in the NM/FSOM junction almost independently of the value of $M_L$.

Thus, we numerically indicated that the charge conductivity in the DSOC is the same as that in the RSOC in the NM/FSOM junction and FM/FSOM junction with $\gamma=z$.
Besides,  at $\alpha = \beta$, the energy dispersion of the DSOC is also the same as that of the RSOC, 
but the spin texture of the DOSC is different from that of the RSOC.
These results implies that the charge conductivity of the FSOM junctions are independent of the spin texture. 
This is main message in this paper, and it will be also discussed by using the transformation in the next section.

\section{Discussion}
\label{sec:III}
From the numerical calculations in the previous section, we will discuss the charge and spin-dependent conductance.
In the previous sections, in order to clarify whether the charge and spin conductance depends on the spin texture of the SOCs, we calculated the transmission probabilities at $\alpha=\beta$, since the energy dispersion of the FSOM is independent of the SOC type but the spin texture depends on the type, RSOC or DSOC (see Fig. \ref{magneticfied}), at $\alpha = \beta$. 
We found that the charge conductance in the RSOC is equal to that in the DSOC [see Eq. (\ref{eq:sum})], and the charge conductance is independent of the spin textures as well as the magnetization $M_R$ in the NM/FSOM junction. 
On the other hand, the spin-dependent conductance $\mathcal{T}^{\gamma, s}_{A}$ depends on the spin textures as well as $M_R$; however there are some relations between $\mathcal{T}^{\gamma, s}_{R}$ and $\mathcal{T}^{\gamma, s}_{D}$.

The relation between $\mathcal{T}^{\gamma, s}_{R}$ and $\mathcal{T}^{\gamma, s}_{D}$ can be understood by the following argument of the transformation. 
At $\alpha =\beta$, the spin texture of the RSOC is changed by that of the DSOC via the spin transformation, $\sigma_x \to \sigma_y$, $\sigma_y \to \sigma_x$, and $\sigma_z \to -\sigma_z$ \cite{Lucignano08,Dario15} , which can be described as $R=i/\sqrt{2}(\sigma_x+\sigma_y)$.
This spin transformation can correspond to the relations between the spin-dependent conductance. 
In $M_R=0$ in the NM/FSOM junction, Eq. (\ref{eq:result3.2}) implied $T^{y,s}_R  \to  T^{x, s}_D $ and $T^{z,\uparrow(\downarrow)}_R  \to  T^{z,\downarrow(\uparrow)}_D $ as the spin transformation $R$ via $\sigma_x \to \sigma_y$, $\sigma_y \to \sigma_x$, and $\sigma_z \to -\sigma_z$. 
In $M_R\neq 0$, The $\mathcal{T}^{\gamma, s}_{R}$ and $\mathcal{T}^{\gamma, s}_{D}$ have the relation as Eq. (11), 
which indicates the transformation $\phi\to-\phi$, $\sigma_x\to-\sigma_y$, $\sigma_y\to\sigma_x$, and $\sigma_z\to\sigma_z$. 
The transformation can be described by $R$ and $P$.
$P$ is the spin and momentum transformation as $k_y\to -k_y$, $\sigma_y\to -\sigma_y$, and $\sigma_z\to-\sigma_z$.
Note that DSOC is even for this transformation $P$.
Applying $P$ after $R$, in $M_R \neq 0$ case, the junction in the RSOC corresponds to that in the DSOC with $\phi\to-\phi$, $\sigma_x\to-\sigma_y$, $\sigma_y\to\sigma_x$, and $\sigma_z\to\sigma_z$; it was shown in Eq. (\ref{eq:result4.2}).
We show that the conductance is invariant under the transformation $R$ and $P$ (see Appendix \ref{app2}).

In the FM/FSOM junction with $M_R=0$, the junction in the RSOC for $\gamma =x$ and $y$ are corresponded to that in the DSOC for $\gamma=y$ and $x$ respectively by the transformation $R$, as shown in $\mathcal{T}^{x, s}_D=\mathcal{T}^{y, s}_R$ in Eq. (\ref{eq:result5.2}).
However, for $\gamma=x$ or $y$, since the FM/FSOM junction in the RSOC does not equal that in the DSOC for same $\gamma$, the charge conductance depended on SOC type and $\gamma$.  
For $\gamma=z$, the junction in the RSOC corresponds to that in the DSOC for $\gamma=z$ by the transformations $R$ and $P$ regardless of $M_R$, as $\mathcal{T}^{z,s}_D = \mathcal{T}^{z, s}_R$ in Eq. (\ref{eq:result5.2}).
Then, the charge conductance is independent of the type of SOC only for $\gamma=z$ in the FM/FSOM junction.

Thus, we gain intuitively understand why the charge conductance is independent of the type of the RSOC and DSOC (or the spin texture of the SOCs) by using spin rotation symmetry.  
These results could imply that the charge conductance is independent of the spin texture rather than depends on the magnitude of the SOC ($\alpha$ and $\beta$).  

\section{Conclusion}
\label{sec:IV}

We have theoretically studied spin-dependent transport in NM/FSOM and FM/FSOM junctions, where the FSOM was applied by an electrical gate tuning the Fermi level, and the FSOM had also RSOC or DSOC with out-of-plane magnetization. 
We have shown the gate voltage dependence of the transmission probability in the RSOC and DSOC under the time-reversal symmetry or time-reversal symmetry breaking system. 
As a result, in the NM/FSOM junction, regardless of the value of $M_R$, 
we have found the relations between the transmission probabilities in the RSOC and that in the DSOC, and the charge conductance is independent of the SOC type.
In the FM/FSOM junction, the gate voltage dependence of the probabilities is similar to that in the NM/FSOM junction, as is the relation between the probabilities in the RSOC and that in the DSOC only when $M_R=0$ or the magnetization in the FM is along the out-of-plane direction.
However, the charge conductance depends on the SOCs and the direction of the magnetization in the FM, unlike that in the NM/FSOM junction.     

In this paper, we compared two systems with the same band structures and different spin textures; i.e., the system in the RSOC and that in the DSOC. Then, several relations between the transmission probabilities in the RSOC and the DSOC were derived.
In particular, it was found that the charge conductance is independent of the type of SOC in some cases. 
This fact shows that in some cases, we can understand the character of the conductance without the SOC details, such as the spin texture.
We expect that this observation will be useful in future studies of spintronics and those involving properties like conductance as explored in our study. 

\section*{Acknowledgment}
This work was supported
by a Grant-in-Aid for Scientific Research on Innovative
Areas, Topological Material Science (Grants No. JP15H05851,
No. JP15H05853 No. JP15K21717), a Grant-in-Aid
for Challenging Exploratory Research (Grant No. JP15K13498)
from the Ministry of Education, Culture, Sports, Science, and
Technology, Japan (MEXT), the Core Research for
Evolutional Science and Technology (CREST) of the Japan
Science and Technology Corporation (JST) (Grant No. JPMJCR14F1).

\appendix
\section{Derivation of the Transmission Probability}\label{app1}
In order to obtain the transmission probability of the NM/FSOM junction under 
the low-temperature limit, we consider the wave function at the Fermi level.
The wave function in NM, $\psi^{z,\uparrow (\downarrow)}(x<0,y)$, is decomposed 
into the injected wave function $\psi^{z,\uparrow(\downarrow)}_{\textrm{in}}$
and the reflected one $\psi^{z,\uparrow(\downarrow)}_{\textrm{ref}}$ as 
\eq{ \label{eq:left}
&\psi^{z,\uparrow (\downarrow)}(x<0,y)=\psi_{\textrm{in}}^{z,\uparrow (\downarrow)}+\psi_{\textrm{ref}}^{z,\uparrow (\downarrow)},\\
&\psi_{\textrm{in}}^{z,\uparrow(\downarrow)}=\chi_{\uparrow(\downarrow)}e^{i(k_x x+k_y y)}\label{eq:in},\\
&\psi_{\textrm{ref}}^{z,\uparrow (\downarrow)}=\left[r^{z,\uparrow(\downarrow)}_\uparrow \chi_\uparrow  +r^{z,\uparrow(\downarrow)}_\downarrow \chi_\downarrow \right]e^{i(-k_x x+k_y y)},
\label{eq:ref}
}
with $\chi_\uparrow = \begin{pmatrix}1 \\ 0 \end{pmatrix}$, $\chi_\downarrow = \begin{pmatrix}0 \\ 1 \end{pmatrix}$, $k_x=k_F\cos\phi$ and $k_y=k_F\sin\phi$.
$\chi_{\uparrow(\downarrow)}$, $r^{z,\uparrow (\downarrow)}_\uparrow$ [$r^{z,\uparrow (\downarrow)}_\downarrow$] is the eigenfunction in NM, the reflection coefficient of up [down] spin electrons with up (down) spin injection. 
Here, we assume that injected electrons are polarized along the $z$ direction.
The transmitted wave function $\psi^{z,\uparrow (\downarrow)}(x>0,y)\equiv\psi_{\textrm{tra}}^{z,\uparrow (\downarrow)}$ is shown as\cite{Oshima18} 
\eq{
&\psi_{\rm tra}^{z,\uparrow (\downarrow)}=t_1^{z,\uparrow (\downarrow)} \chi_1(\bm{k}_1)e^{i\bm{k}_1\cdot\bm{r}}+t_2^{z,\uparrow (\downarrow)}\chi_2(\bm{k}_2)e^{i\bm{k}_2\cdot\bm{r}},
\label{transmissionwave}
}
with
\begin{align}
&\chi_1(\bm{k})=\theta(\Delta)\chi_+(\bm{k})+\theta(-\Delta)\chi_-(\bm{k}),\quad \Delta\equiv E-eV_g+E_c,  \\
&\chi_2(\bm{k})=\chi_-(\bm{k}),\\
&\chi_+(\bm{k})=\begin{pmatrix} g_+(\bm{k}) \\ 1 \end{pmatrix},\quad \chi_-(\bm{k})=\begin{pmatrix} 1 \\  g_-(\bm{k}) \end{pmatrix}.
\end{align}
Here, $t_1^{z,\uparrow (\downarrow)}$ $[t_2^{z,\uparrow (\downarrow)}]$ denotes the transmission coefficient with up (down) spin injection. 
$\chi_\pm$ is the eigenfunction for the eigenvalue in FSOM.
$\bm{k}_1 = (k_{1,x},k_y )$ and $\bm{k}_2 = (k_{2,x},k_y )$ are the momentum in FSOM, which are defined for $k_1^2\leq k_2^2$ with $k_{1(2)}^2=k_{1(2),x}^2+k_y^2$. 
We set $g_\pm$ as follows:
\begin{align}
g_\pm(\bm{k})=\left\{
\begin{aligned}
&\frac{-\alpha i\left(k_x\mp ik_y\right)}{M_R+\sqrt{ \alpha^2 k^2+M_R^2}} \qquad(\textrm{RSOC}),\\
&\frac{\beta\left(\pm k_x+ ik_y\right)}{M_R+\sqrt{ \beta^2 k^2+M_R^2}} \qquad(\textrm{DSOC}).
\end{aligned}
\right.
\end{align}

From the energy dispersion of FSOM, $k_{1(2)}^2$ is given by
\begin{align}
k_{1(2)}^2&=\frac{2m}{\hbar^2}\left[(E_F-eV_g)+2E_{\rm SOC}\right.\nonumber\\
&\qquad\qquad\left.-(+)\sqrt{4E_{\rm SOC} (E_F-eV_g)+4E_{\rm SOC}^2+M_R^2}\right].
\label{k1(2)}
\end{align}
We define $E_{\rm SOC}=m \alpha^2/(2\hbar^2)$ or $m \beta^2/(2\hbar^2)$ corresponding to the SOCs.
The signs of $k_{1,x}$ and $k_{2,x}$ are determined so that the velocity $v_x$ takes a positive value, because the electron is injected along the $x$ direction.
The velocity operator $v_x=\partial H/(\hbar\partial k_x)$ is given by Eq. (\ref{eq:total}) as \cite{Streda03,Srisongmuang08,Dario15} 
\eq{
v_x =\left\{
\begin{aligned}
&\frac{\hbar k_x }{m} +\frac{\alpha}{\hbar} \theta(x) \sigma_y\qquad(\textrm{RSOC}),\\
&\frac{\hbar k_x }{m} +\frac{\beta}{\hbar} \theta(x) \sigma_x \qquad(\textrm{DSOC}).
\end{aligned}
\right. 
\label{velocity}
}
When $k_{1,x}$ ($k_{2,x}$) becomes an imaginary number; its sign is determined so that $\chi_{\pm} \to 0$ in the limit of $x \to \infty$.

The boundary condition at the interface \cite{Molenkamp01,Srisongmuang08,Zulicke01,Jantayod13,Fukumoto15,Reeg17} is given as follows:
\begin{align}
\begin{aligned}
&\psi^{z,\uparrow (\downarrow)}(+0,y)-\psi^{z,\uparrow (\downarrow)}(-0,y)=0,\\
&v_x[\psi^{z,\uparrow (\downarrow)}(+0,y)-\psi^{z,\uparrow (\downarrow)}(-0,y)]=\frac{2U}{i\hbar}\psi^{z,\uparrow (\downarrow)}(0,y).
\end{aligned}
\label{eq:condition}
\end{align}
From this condition, we have a equation about the coefficients as follow:
\begin{align}
&\begin{pmatrix}\chi_1 & \chi_2 & -\chi_\uparrow & -\chi_\downarrow \\ 
\left(v_x-\frac{2U}{i\hbar}\right)\chi_1 & \left(v_x-\frac{2U}{i\hbar}\right)\chi_2 & -v_x\chi_\uparrow & -v_x\chi_\downarrow\end{pmatrix}
\begin{pmatrix}t_1^{z,\uparrow (\downarrow)}\\ t_2^{z,\uparrow (\downarrow)}\\ r_\uparrow^{z,\uparrow (\downarrow)}\\ r_\downarrow^{z,\uparrow (\downarrow)}\end{pmatrix}\nonumber\\
&\qquad\qquad=
\begin{pmatrix}\chi_{\uparrow(\downarrow)} \\ v_x\chi_{\uparrow(\downarrow)}\end{pmatrix}.
\label{eq:coefficient}
\end{align} 
Solving the coefficient from this equation, we obtain the transmission probability at each angle $T^{z,\uparrow (\downarrow)}(\phi)$ as\cite{Zulicke01,Srisongmuang08,Jantayod13,Oshima18} 
\begin{align}
T^{z,\uparrow (\downarrow)}(\phi)&={\rm Re}\left|\frac{\psi_{\rm tra}^{z,\uparrow (\downarrow)\dagger} v_x\psi^{z,\uparrow (\downarrow)}_{\rm tra}}{\psi_{\rm in}^{z,\uparrow (\downarrow)\dagger} v_x\psi^{z,\uparrow (\downarrow)}_{\rm in}}\right| =1-{\rm Re}\left|\frac{\psi_{\rm ref}^{z,\uparrow (\downarrow)\dagger} v_x\psi^{z,\uparrow (\downarrow)}_{\rm ref}}{\psi_{\rm in}^{z,\uparrow (\downarrow)\dagger} v_x\psi^{z,\uparrow (\downarrow)}_{\rm in}}\right| \nonumber\\
                                           &=1-\left({|r^{z,\uparrow (\downarrow)}_\uparrow|}^2+{|r^{z,\uparrow (\downarrow)}_\downarrow|}^2\right).
\label{T1}
\end{align}
Using the Landauer formula, we describe an electric current between two leads, $I$, as follows \cite{Sablikov07,Srisongmuang08,Jantayod13,Oshima18}:
\begin{align}
I= \frac{e^2VL}{4\pi^2\hbar}\int_{-\pi/2}^{\pi/2} d\phi \cos\phi\cdot k_F[T^{z,\uparrow}(\phi)+T^{z,\downarrow}(\phi)].
\label{current}
\end{align}
The first term in Eq. (\ref{current}) is proportional to $\mathcal{T}^{z,\uparrow}$, and the second term is $\mathcal{T}^{z,\downarrow}$.
Thus, $\mathcal{T}^{\gamma,\uparrow} + \mathcal{T}^{\gamma,\downarrow}$ is proportional to the conductance, because the conductance is proportional to the current at the low-temperature limit.

\section{Spin-dependent Conductance under the Unitary Transformation}\label{app2}
In this appendix, we will show that the conductance are invariant under an unitary transformation $U$, which is assumed $k_x\overset{U}{\to} k_x$ and $\partial U/\partial k_x =0$.
Applying $U$, the Hamiltonian of the junction [in Eqs. (\ref{eq:total}) and (\ref{eq:FMFSOM})] and the eigenfunctions in the each side are changed as $H\to \acute{H}=UHU^\dagger$ and $\chi_{\uparrow(\downarrow,1,2)}\to\acute{\chi}_{\uparrow(\downarrow,1,2)}=U\chi_{\uparrow(\downarrow,1,2)}$.  
Then,  the velocity operator in $\acute{H}$ becomes
\begin{align}
\acute{v}_x=Uv_xU^\dagger .
\end{align}  
From the boundary condition in Eq. (\ref{eq:condition}) under $U$, the transmission and reflection coefficients, $\acute{t}^{z,\uparrow(\downarrow)}_{1[2]}$ and $\acute{r}^{z,\uparrow(\downarrow)}_{\uparrow[\downarrow]}$ are given by:
\begin{align}
&\begin{pmatrix}U&0 \\ 0&U\end{pmatrix}\begin{pmatrix}\chi_1 & \chi_2 & -\chi_\uparrow & -\chi_\downarrow \\ 
\left(v_x-\frac{2U}{i\hbar}\right)\chi_1 & \left(v_x-\frac{2U}{i\hbar}\right)\chi_2 & -v_x\chi_\uparrow & -v_x\chi_\downarrow\end{pmatrix}\nonumber\\
&\qquad\cdot\begin{pmatrix}\acute{t}_1^{z,\uparrow (\downarrow)}\\ \acute{t}_2^{z,\uparrow (\downarrow)}\\ \acute{r}_\uparrow^{z,\uparrow (\downarrow)}\\ \acute{r}_\downarrow^{z,\uparrow (\downarrow)}\end{pmatrix}
=
\begin{pmatrix}U&0 \\ 0&U\end{pmatrix}\begin{pmatrix}\chi_{\uparrow(\downarrow)} \\ v_x\chi_{\uparrow(\downarrow)}\end{pmatrix}.
\label{eq:coefficientdash}
\end{align} 
Note that $U$ is a $2\times 2$ matrix and $\chi_{\uparrow(\downarrow,1,2)}$ are two component vectors. 
Here, we notice that  Eq. (\ref{eq:coefficientdash}) is the same as Eq. (\ref{eq:coefficient}), 
i.e., $\acute{t}^{z,\uparrow(\downarrow)}_{1[2]}=t^{z,\uparrow(\downarrow)}_{1[2]}$ and $\acute{r}^{z,\uparrow(\downarrow)}_{\uparrow[\downarrow]}=r^{z,\uparrow(\downarrow)}_{\uparrow[\downarrow]}$.
Then, we obtain $\acute{\psi}^{z,\uparrow(\downarrow)}_{\rm in[ref,tra]}=U\psi^{z,\uparrow(\downarrow)}_{\rm in[ref,tra]}$, where $\acute{\psi}^{z,\uparrow(\downarrow)}_{\rm in[ref,tra]}$ is the wave function after the transformation.
As a result, we find that the transmission probabilities are invariant about the transformation $U$:
\begin{align}
{\rm Re}\left|\frac{\psi_{\rm tra}^{z,\uparrow (\downarrow)\dagger} v_x\psi^{z,\uparrow (\downarrow)}_{\rm tra}}{\psi_{\rm in}^{z,\uparrow (\downarrow)\dagger} v_x\psi^{z,\uparrow (\downarrow)}_{\rm in}}\right|
&={\rm Re}\left|\frac{\acute{\psi}_{\rm tra}^{z,\uparrow (\downarrow)\dagger}U v_x U^\dagger \acute{\psi}^{z,\uparrow (\downarrow)}_{\rm tra}}{\acute{\psi}_{\rm in}^{z,\uparrow (\downarrow)\dagger}U v_x U^\dagger \acute{\psi}^{z,\uparrow (\downarrow)}_{\rm in}}\right|\nonumber\\
&={\rm Re}\left|\frac{\acute{\psi}_{\rm tra}^{z,\uparrow (\downarrow)\dagger}\acute{v}_x \acute{\psi}^{z,\uparrow (\downarrow)}_{\rm tra}}{\acute{\psi}_{\rm in}^{z,\uparrow (\downarrow)\dagger}\acute{v}_x \acute{\psi}^{z,\uparrow (\downarrow)}_{\rm in}}\right|.
\end{align}
Note that transformations $R$ and $P$ are satisfied the assumption of $U$.

\end{document}